\documentclass[english]{article}
\usepackage[T1]{fontenc}
\usepackage[latin9]{inputenc}
\usepackage{float}
\usepackage{amsmath}
\usepackage{amssymb}
\usepackage{graphicx}

\makeatletter

\providecommand{\tabularnewline}{\\}

\newcommand{\lyxaddress}[1]{
\par {\raggedright #1
\vspace{1.4em}
\noindent\par}
}

\makeatother

\usepackage{babel}
\begin{document}

\title{Calibration Of Proton Accelerator Beam Energy}

\author{D. Cohen, M. Friedman, M. Paul}
\maketitle

\lyxaddress{Racah Institute of Physics, The Hebrew University, Jerusalem, Israel,
91904}
\begin{abstract}
When studying the $^{7}Li\left(p,n\right)^{7}Be$ reaction with a
RF accelerator, it is difficult to define the precise energy of the
beam and its energy distribution, which together fully define the
beam, for our purpose. What we do know is the difference between two
given energies. We resolve this problem by finding a reference energy
and then finding another energy by examining the data, relative to
the reference energy. From the difference between them we can approximate
the energy distribution for a given energy, and since we know the
energy threshold of the reaction ($E_{threshold}$ for $^{7}Li\left(p,n\right)^{7}Be$
is about $1880.4-1880.8\:(keV)$) we can calibrate the beam energy.
We then determine the energy which produce the maximum yield derivative
as the reference and record the energy whose neutron yield is 5\%
of the reference energy. The data is collected by simulating this
reaction using SimLit \cite{simlit}. Since this is a simulation we
know the real energy distribution so we make a linear fit for energy
distribution as function of energy difference. We tested the theory
on experimental data for which we approximate the energy distribution
by other means, and found our new method to be accurate and satisfactory
for our needs.
\end{abstract}

\section{introduction}

Our research uses a RF proton accelerator to produce $^{7}Li\left(p,n\right)^{7}Be$,
and it is crucial to define the beam energy as accurately as we can.
To do so we need the exact energy value and the energy spread (standard
deviation), that from now on will be referred to as ``energy sigma''.
Although we don't know the two values mentioned above, we do know
the gap between any two energies the accelerator produces. Moreover,
we can measure the number of neutron yield for a given angle with
a Long Counter Detector. To solve our problem, we search for a well
defined energy as a reference. By extracting this energy we can find
another energy that conserves some relation value which we explain
later on. After collecting these differences and their energy sigma
we assume they fit each other linearly. Moreover we assume that the
energy differences can point out the real value of the reference energy
again by linear fit. This method is a new way to calibrate the proton
beam energy. By putting in the equations the measured value of the
energy difference extracted from a given experimental data, we can
get the energy sigma and the real value of the reference energy (``new
reference energy''). The rest is calibrated in reference to the new
reference energy, since as we mentioned before we do know the gap
between any two given energies. We tested this method and confirmed
it on experimental data. Furthermore, we found that the angle contributions\footnote{Meaning we made the fit of energy sigma and reference energy value
vs. energy difference for different angles and the fits were almost
the same.} for this method in a Long Counter Detector are negligible.

\section{methods}

\subsection{simulation with LiLiT}

\begin{figure}[H]
\begin{centering}
\includegraphics[width=14cm]{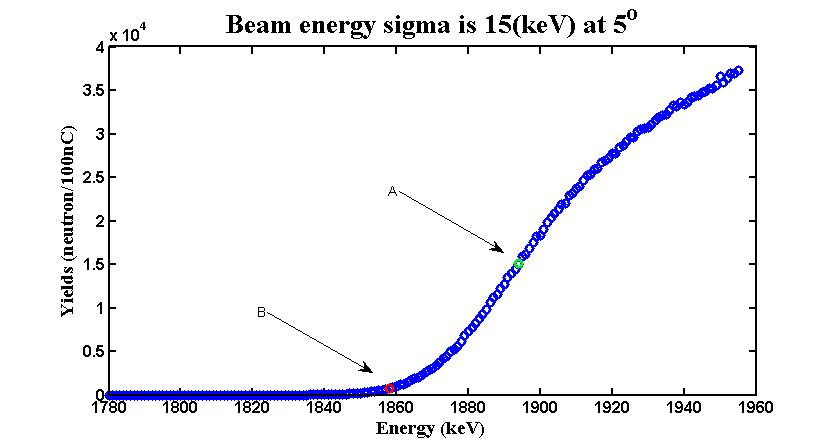}
\par\end{centering}
\begin{centering}
\includegraphics[width=14cm]{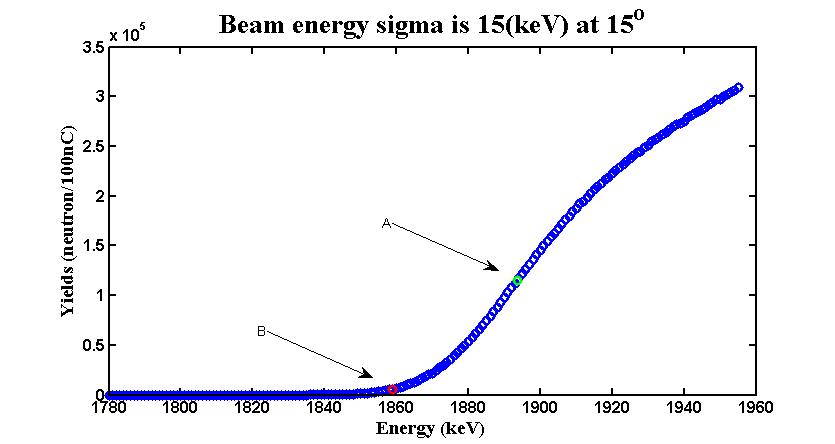}
\par\end{centering}
\caption{\label{fig:neutron Vs energy}Example of neutron yield vs. energy.
As plotted by SimLiT simulation at two differently angles - 5 and
15 degrees, between the direction of proton beam and the scattered
neutron direction and the same energy spread. Point A in the graph
is the reference energy as we extract from the derivative (shown in
fig \eqref{fig:neutron Vs energy derive}) and point B is the energy
that have 5\% of the yield of point A. As can be observed the two
graphs are almost similar, suggesting the that the angle dependent
is negligible. Confirmed by many different sets of data.}
\end{figure}

\begin{figure}[H]
\begin{centering}
\includegraphics[width=14cm]{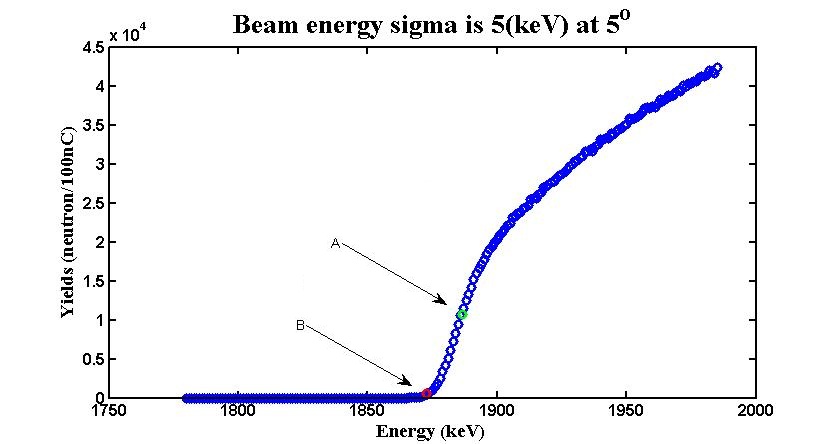}
\par\end{centering}
\centering{}\includegraphics[width=14cm]{15kev_y_5d.jpg}\caption{\label{fig:neutron Vs energy 1}Example of neutron yield vs. energy.
As plotted by SimLiT simulation at two differently energies spread
- 5 and 15 keV and the same angle. Point A in the graph is the reference
energy as we extract from the derivative (shown in fig \eqref{fig:neutron Vs energy derive 1})
and point B is the energy that have 5\% of the yield of point A. As
can be observed the two graphs are different. The reference energy
shifted to an higher energy and the energy difference between the
the two energies as grow bigger. Suggesting the that the energy spread
and the real value of the reference energy are a characteristics variables
of the problem. Confirmed by many different sets of data.}
\end{figure}

First we did simulation with SimLit at wide spectrum in terms of energies,
angles\footnote{The angle between the direction of the proton beam to the scattered
neutron.}, and energy sigma. Of $^{7}Li\left(p,n\right)^{7}Be$ reaction caused
by accelerate protons on lithium target. The simulation sets we did
where to take a constant energy spread and constant maximum angle,
meaning we count all neutrons who scattered from the center of the
target up to the maximum angle we determine. And we let the simulation
to run on a rang of energies far less then the reaction threshold
and up to 2 MeV. The simulation based on ``Monte Carlo'' that rely
on repeated random sampling to compute their results. Verifying we
make at least 10,000 events compensate the statistical error which
is square rout of the samples number and to be as accurate as 1\%.
The data we extract were the excitation function and they look like
figs \eqref{fig:neutron Vs energy} and \eqref{fig:neutron Vs energy 1}.
 
\subsection{analysis of computed data}

\begin{figure}[H]
\begin{centering}
\includegraphics[width=14cm]{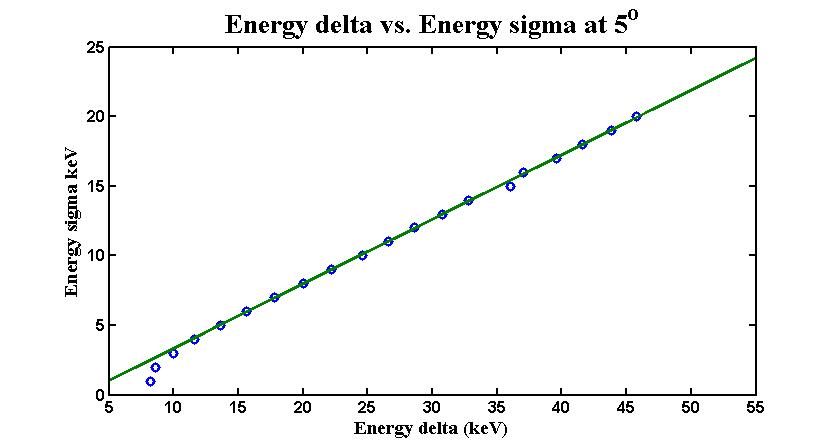}
\par\end{centering}
\begin{centering}
\includegraphics[width=14cm]{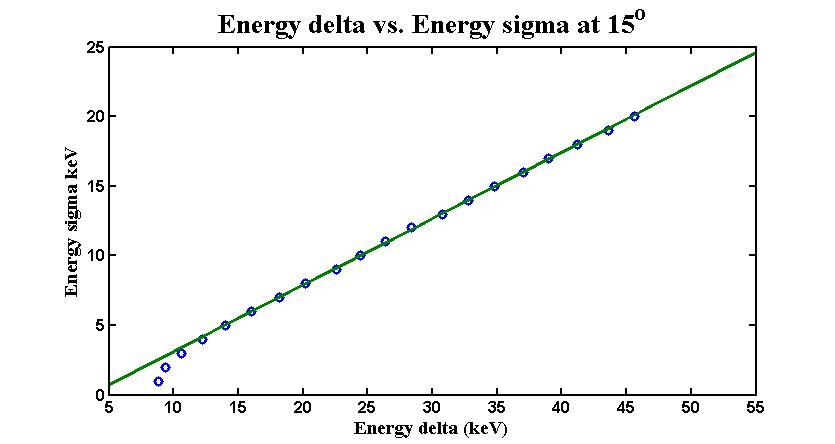}
\par\end{centering}
\caption{\label{fig:delta vs sigma}The graph represent the data of the energy
gap between the reference energy and the low energy vs the known energy
spread. As extract from SimLiT simulation. The grin graph is the linear
fit we suggested. As can be seen our fit is not accurate at small
values of energy gap, smaller than 10 keV. At thous energy gaps we
can estimate by other means and we do not relay on our method. In
contrary at high energy spread our method is fit very well and it
can use as analysis tool. Moreover we can see that the two fits at
different angles aren't differ from each other. Confirming that the
angle dependent is negligible.}
\end{figure}

\begin{figure}[H]
\begin{centering}
\includegraphics[width=14cm]{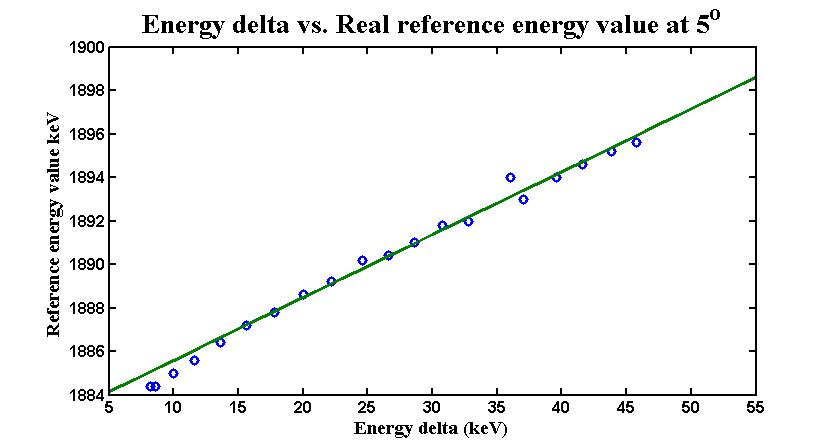}
\par\end{centering}
\begin{centering}
\includegraphics[width=14cm]{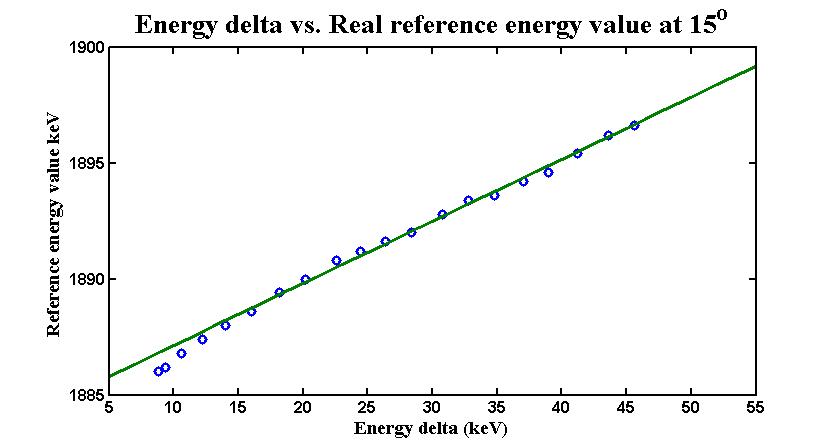}
\par\end{centering}
\caption{\label{fig:delta vs real value}The graph represent the data of the
energy gap between the reference energy and the low energy vs the
known real value of the reference energy. As extract from SimLiT simulation.
The grin graph is the linear fit we suggested. As can be seen our
fit is not accurate at small values of energy gap, smaller than 15
keV. At thous energy gaps we can estimate by other means and we do
not relay on our method. In contrary at high energy spread our method
is fit very well and it can use as analysis tool. Moreover we can
see that the two fits at different angles aren't differ from each
other. Confirming that the angle dependent is negligible.}
\end{figure}

To find a uniquely defined energy in the excitation function (shown
in fig \eqref{fig:neutron Vs energy}) we take the derivative of the
this function\footnote{The excitation function as bean smooth by interpolation fit and putting
many extra energies. The derivative as bean fitted to an superposition
of two gaussian to avoid noise.} (shown in fig \eqref{fig:neutron Vs energy}) and as can be seen
we found a very distinct peak of yield density \footnote{yield density is define to be the derivative of the yield vs. energy
function.}. We chose the energy that fits this peak as our reference energy.
We define the lower energy as the energy which produces around 5\%
of the reference energy neutron yield. This act prevents noise interference.
After doing so for several simulations we extract the values of the
difference between the two energies in each experiment, and we know
the energy sigma and the real value of the reference energy because
this was a simulation. We did this analysis for different sets with
the same statistics per cross section. From this data we extract a
linear fit of energy sigma as a function of the gap, fig \eqref{fig:delta vs sigma}.
And a linear fit of reference energy value as a function of the gap,
fig \eqref{fig:delta vs real value}.
\begin{figure}[H]
\begin{centering}
\includegraphics[width=14cm]{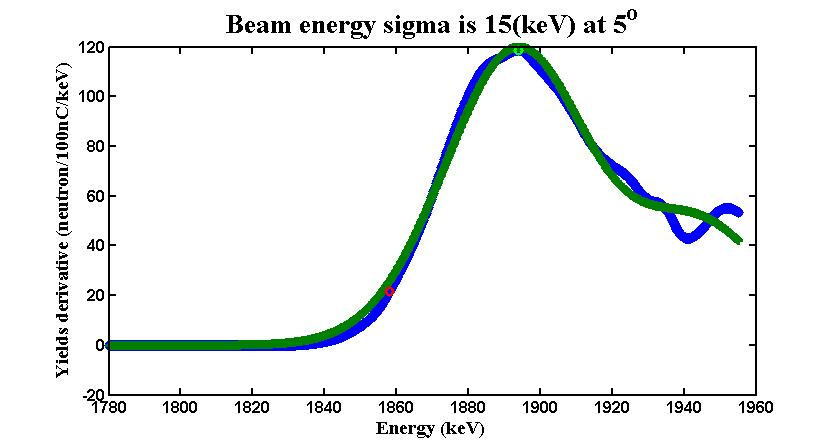}
\par\end{centering}
\begin{centering}
\includegraphics[width=14cm]{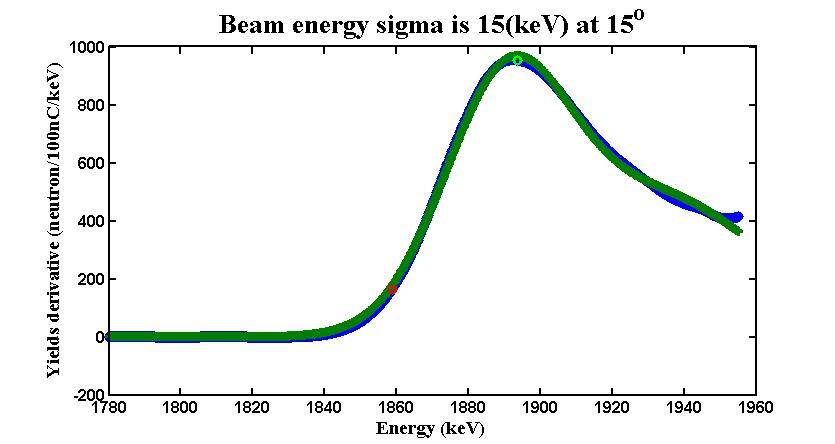}
\par\end{centering}
\caption{\label{fig:neutron Vs energy derive}Example of the derivative of
the excitation functions, fig \eqref{fig:neutron Vs energy}. The
blue graphs is the actual derivation and the green graph is fit of
superposition of two gaussian, which give the best results for this
method. The reference energy is the global maximum of the fit. As
can be observed the two graphs are almost similar, suggesting the
that the angle dependent is negligible. Confirmed by many different
sets of data.}
\end{figure}
\begin{figure}
\begin{centering}
\includegraphics[width=14cm]{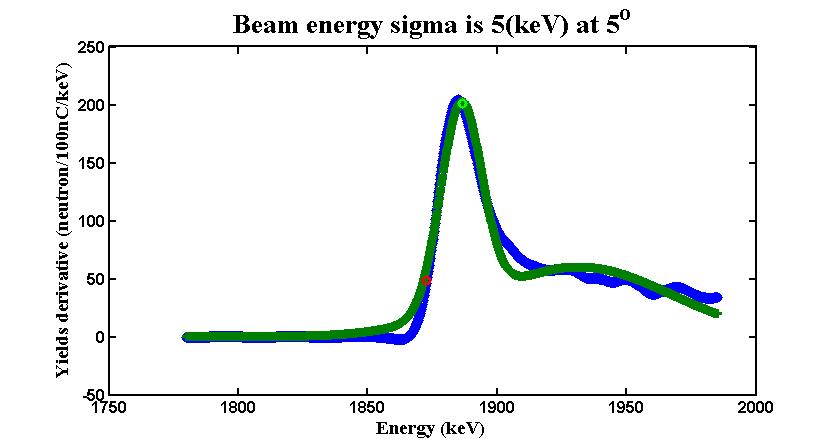}
\par\end{centering}
\begin{centering}
\includegraphics[width=14cm]{15kev_yd_5d.jpg}
\par\end{centering}
\caption{\label{fig:neutron Vs energy derive 1}Example of the derivative of
the excitation functions, fig \eqref{fig:neutron Vs energy 1}. The
blue graphs is the actual derivation and the green graph is fit of
superposition of two gaussian, which give the best results for this
method. The reference energy is the global maximum of the fit. As
can be observed the two graphs are different. The reference energy
shifted to an higher energy and the energy difference between the
the two energies as grow bigger. Suggesting the that the energy spread
and the real value of the reference energy are a characteristics variables
of the problem. Confirmed by many different sets of data.}
\end{figure}

\subsection{the data from the simulation}

\begin{table}[H]
\begin{centering}
\begin{tabular}{|c|c|c|c|c|}
\hline 
index & energy sigma & low energy & reference energy & energy difference\tabularnewline
\hline 
\hline 
1 & 1 & 1876.2 & 1884.4 & 8.2\tabularnewline
\hline 
2 & 2 & 1875.8 & 1884.4 & 8.6\tabularnewline
\hline 
3 & 3 & 1875 & 1885 & 10\tabularnewline
\hline 
4 & 4 & 1874 & 1885.6 & 11.6\tabularnewline
\hline 
5 & 5 & 1872.8 & 1886.4 & 13.6\tabularnewline
\hline 
6 & 6 & 1871.6 & 1887.2 & 15.6\tabularnewline
\hline 
7 & 7 & 1870 & 1887.8 & 17.8\tabularnewline
\hline 
8 & 8 & 1868.6 & 1888.6 & 20\tabularnewline
\hline 
9 & 9 & 1867 & 1889.2 & 22.2\tabularnewline
\hline 
10 & 10 & 1865.6 & 1890.2 & 24.6\tabularnewline
\hline 
11 & 11 & 1863.8 & 1890.4 & 26.6\tabularnewline
\hline 
12 & 12 & 1862 & 1891 & 28.6\tabularnewline
\hline 
13 & 13 & 1861 & 1891.8 & 30.8\tabularnewline
\hline 
14 & 14 & 1859.2 & 1892 & 32.8\tabularnewline
\hline 
15 & 15 & 1858 & 1894 & 36\tabularnewline
\hline 
16 & 16 & 1856 & 1893 & 37\tabularnewline
\hline 
17 & 17 & 1854.4 & 1894 & 39.6\tabularnewline
\hline 
18 & 18 & 1853 & 1894.6 & 41.6\tabularnewline
\hline 
19 & 19 & 1851.4 & 1895.2 & 43.8\tabularnewline
\hline 
20 & 20 & 1849.8 & 1895.6 & 45.8\tabularnewline
\hline 
\end{tabular}
\par\end{centering}
\caption{\label{tab:The-data-from}The data from SimLiT simulation at 5 degrees.
Energy's value presented in keV. The data ploted on figs \eqref{fig:delta vs sigma},\eqref{fig:delta vs real value}.}
\end{table}
\begin{table}[H]
\begin{centering}
\begin{tabular}{|c|c|c|c|c|}
\hline 
index & energy sigma & low energy & reference energy & energy difference\tabularnewline
\hline 
\hline 
1 & 1 & 1877.2 & 1886 & 8.8\tabularnewline
\hline 
2 & 2 & 1876.8 & 1886.2 & 9.4\tabularnewline
\hline 
3 & 3 & 1876.2 & 1886.8 & 10.6\tabularnewline
\hline 
4 & 4 & 1875.2 & 1887.4 & 12.2\tabularnewline
\hline 
5 & 5 & 1874 & 1888 & 14\tabularnewline
\hline 
6 & 6 & 1872.6 & 1888.6 & 16\tabularnewline
\hline 
7 & 7 & 1871.2 & 1889.4 & 18.2\tabularnewline
\hline 
8 & 8 & 1869.8 & 1890 & 20.2\tabularnewline
\hline 
9 & 9 & 1868.2 & 1890.8 & 22.6\tabularnewline
\hline 
10 & 10 & 1866.8 & 1891.2 & 24.4\tabularnewline
\hline 
11 & 11 & 1865.2 & 1891.6 & 26.4\tabularnewline
\hline 
12 & 12 & 1863.6 & 1892 & 28.4\tabularnewline
\hline 
13 & 13 & 1862 & 1892.8 & 30.8\tabularnewline
\hline 
14 & 14 & 1860.6 & 1893.4 & 32.8\tabularnewline
\hline 
15 & 15 & 1858.8 & 1893.6 & 34.8\tabularnewline
\hline 
16 & 16 & 1857.2 & 1894.2 & 37\tabularnewline
\hline 
17 & 17 & 1855.6 & 1894.6 & 39\tabularnewline
\hline 
18 & 18 & 1854.2 & 1895.4 & 41.2\tabularnewline
\hline 
19 & 19 & 1852.6 & 1896.2 & 43.6\tabularnewline
\hline 
20 & 20 & 1851 & 1896.6 & 45.6\tabularnewline
\hline 
\end{tabular}
\par\end{centering}
\caption{\label{tab:The-data-from-2}The data from SimLiT simulation at 15
degrees. Energy's value presented in keV. The data ploted on figs
\eqref{fig:delta vs sigma},\eqref{fig:delta vs real value}.}
\end{table}

Here is the function that we extracted : 
\begin{equation}
\sigma(\vartriangle E)=0.463\bullet\vartriangle E-1.3\label{eq:general}
\end{equation}
\begin{equation}
E_{reference}(\bigtriangleup E)=0.289\bullet\bigtriangleup E+1883\label{eq:calibration}
\end{equation}
And this is plot of the data with their fit. 

\subsection{using those fits to calibrate experimental data}

Assuming this phenomena is an continues phenomena. We can imagine
that if we do the same analysis on a real experimental data, who done
approximately at the same conditions. We will get a value for the
gap between the high energy and the low one. Now instead of making
a set of points we gust put the energy gap in the fit we gust mentioned,
and extract the energy spread an the real absolute value of the reference
energy. This method doesn't take in to account any physical dimensions
of the problem, and it doesn't use any absolute value of the problem
but the peak at the derivative function which is a characteristic
of the problem.

\section{results}

Our results as we extracted from the data by these means are shown
here, we present three different data sets, that we estimated the
energy sigma and the shift of the energy axis, using our new method
and other methods\footnote{TRIM calculation and Geant4 simulation}
for cooperation.

First data set we approximate sigma to be 1.5keV while the new method
shows. 
\begin{equation}
\sigma(\vartriangle E=2.6\:(keV))=0.463\bullet2.6-1.3=-0.096\:(keV)\label{eq:sigma 1}
\end{equation}
\begin{equation}
E_{reference}(\bigtriangleup E=2.6\:(keV))=0.289\bullet2.6+1883=1883.4\:(keV)\label{eq:calibration-1}
\end{equation}
\begin{figure}[H]
\includegraphics[width=14cm]{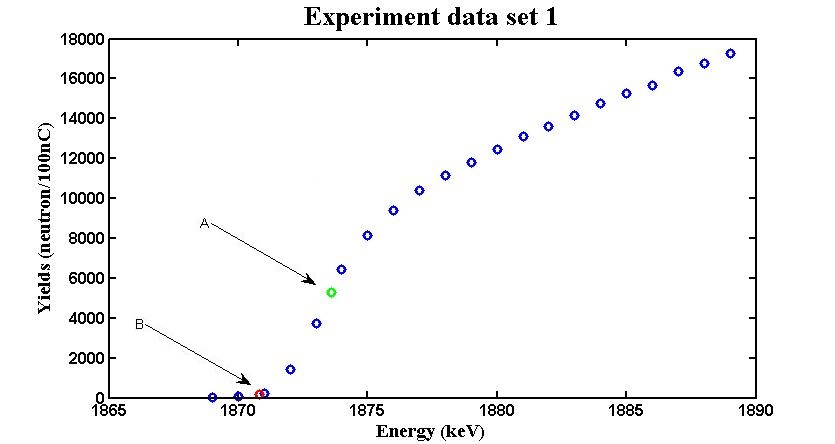}

\includegraphics[width=14cm]{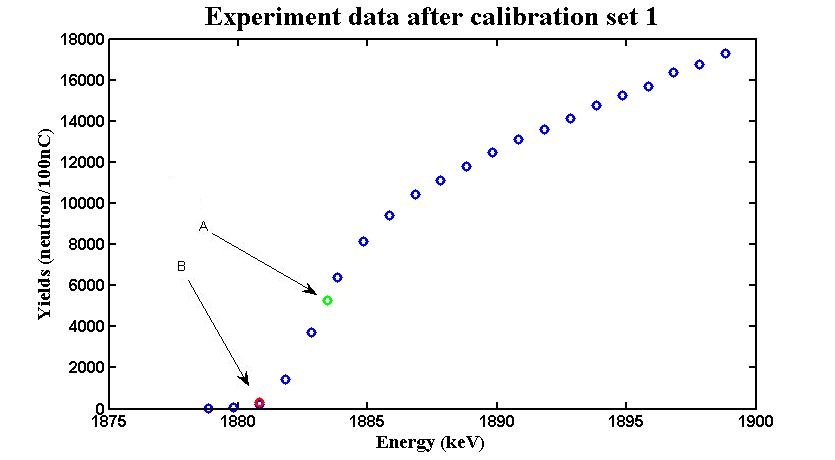}

\includegraphics[width=14cm]{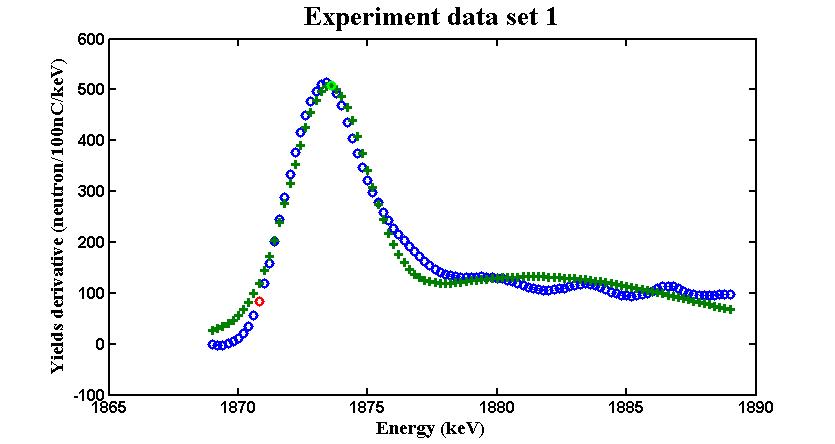}\caption{\label{fig:Neutron-yields-Vs 1-1}Example of real experiment analysis
of data set \cite{key-1} with narrow energy sigma. Point A in the
graph is the reference energy as we extract from the derivative (third
graph in this data set) and point B is the energy that have 5\% of
the yield of point A. The first graph is the yield of neutrons counted
by long counter detector time normalized, and the energies is the
energies the accelerator measured before it hit the target. The second
graph is the same as the first one after calibration of the energy
axis with our new method. At the third graph the blue graphs is the
actual derivation and the green graph is fit of superposition of two
gaussian, which give the best results for this method. The reference
energy is the global maximum of the fit. Exact values can be seen
at table \eqref{tab:real data}.}
\end{figure}

Second data set our approximation sigma is 15keV while the new method
shows. 
\begin{equation}
\sigma(\vartriangle E=55.4\:(keV))=0.463\bullet55.4-1.3=24.37\:(keV)\label{eq:sigma 2}
\end{equation}
\begin{equation}
E_{reference}(\bigtriangleup E=55.4\:(keV))=0.289\bullet55.4+1883=1898.7\:(keV)\label{eq:calibration-2}
\end{equation}
\begin{figure}[H]
\begin{centering}
\includegraphics[width=14cm]{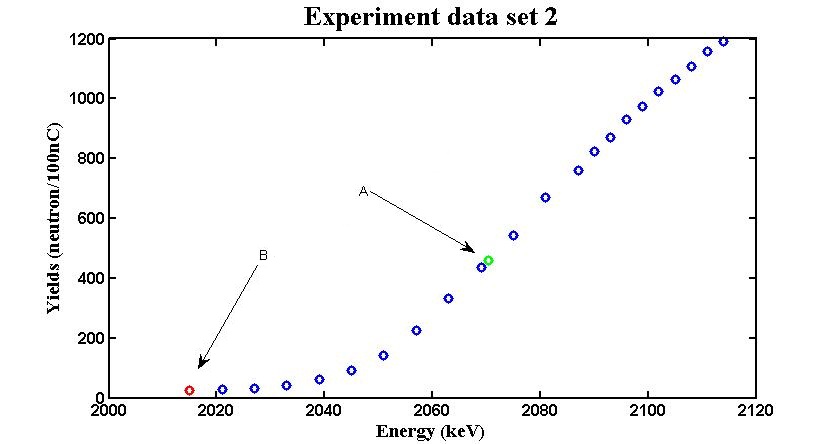}
\par\end{centering}
\begin{centering}
\includegraphics[width=14cm]{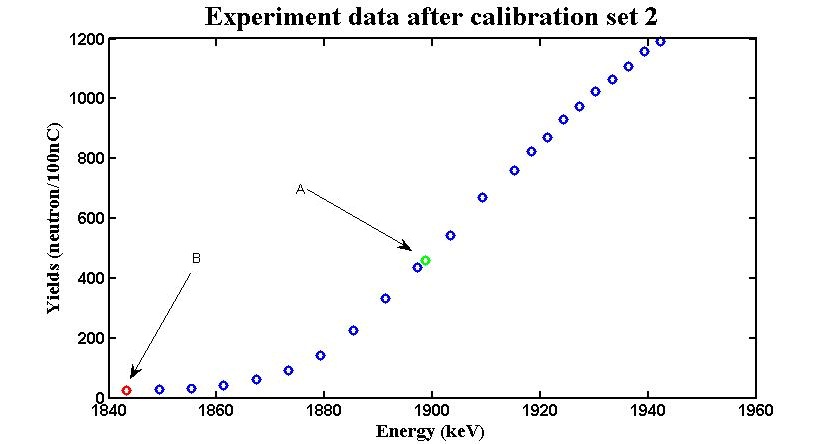}
\par\end{centering}
\centering{}\includegraphics[width=14cm]{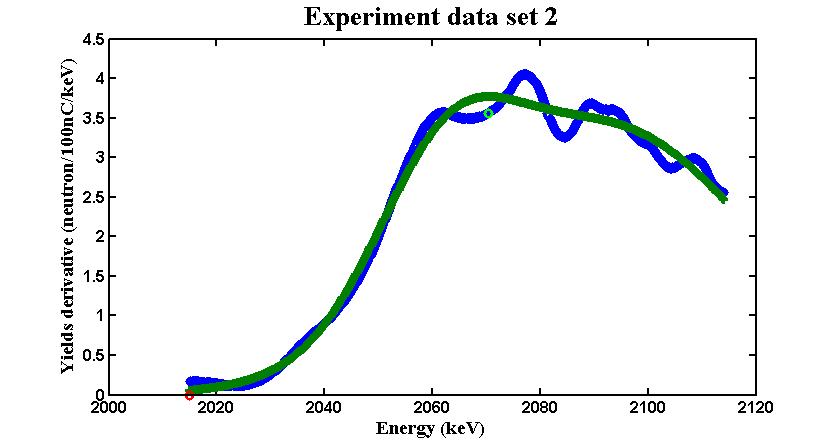}\caption{\label{fig:Neutron-yields-Vs 1-2}Example of real experiment analysis
of data set \cite{key-2} with wide energy sigma. Point A in the graph
is the reference energy as we extract from the derivative (third graph
in this data set) and point B is the energy that have 5\% of the yield
of point A. The first graph is the yield of neutrons counted by long
counter detector time normalized, and the energies is the energies
the accelerator measured before it hit the gold foil, which spread
the energy to a wide spectra. The second graph is the same as the
first one after calibration of the energy axis with our new method.
At the third graph the blue graphs is the actual derivation and the
green graph is fit of superposition of two gaussian, which give the
best results for this method. The reference energy is the global maximum
of the fit. Exact values can be seen at table \eqref{tab:real data}.}
\end{figure}

Third data set our approximation sigma is 20keV and shift of keV while
the new method shows.
\begin{equation}
\sigma(\vartriangle E=53.2\:(keV))=0.463\bullet53.2-1.3=23.35\:(keV)\label{eq:sigma 3}
\end{equation}
\begin{equation}
E_{reference}(\bigtriangleup E=53.2\:(keV))=0.289\bullet53.2+1883=1898.1\:(keV)\label{eq:calibration-3}
\end{equation}
\begin{figure}[H]
\begin{centering}
\includegraphics[width=14cm]{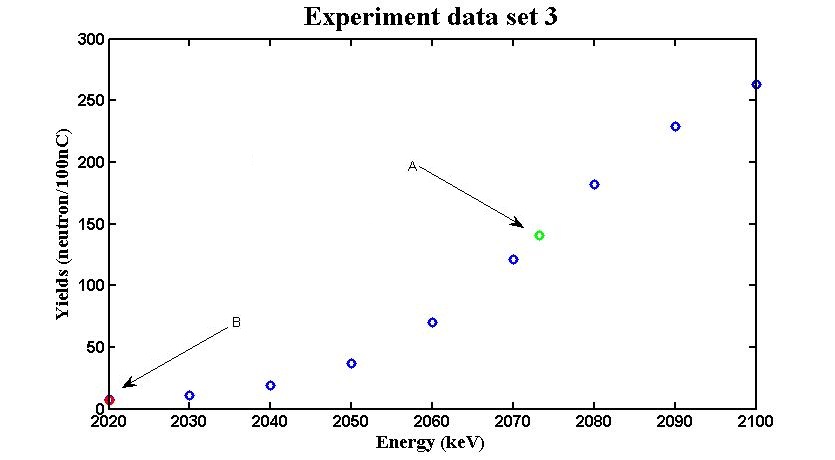}
\par\end{centering}
\begin{centering}
\includegraphics[width=14cm]{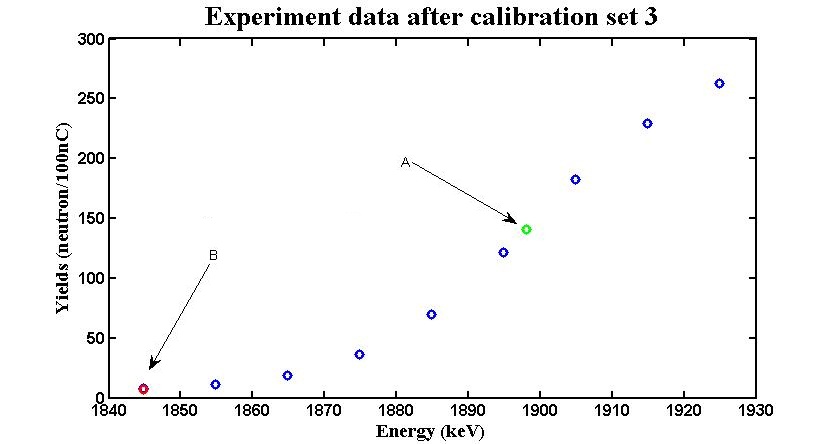}
\par\end{centering}
\centering{}\includegraphics[width=14cm]{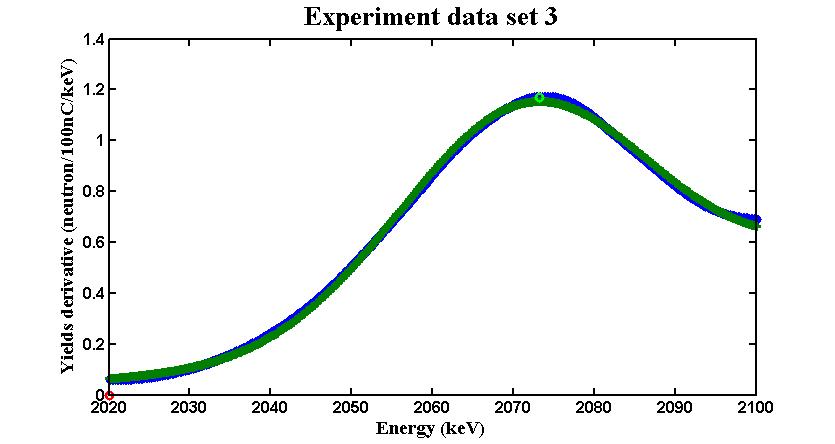}\caption{\label{fig:Neutron-yields-Vs 2-1}Example of real experiment analysis
of data set \cite{key-3} with wide energy sigma. Point A in the graph
is the reference energy as we extract from the derivative (third graph
in this data set) and point B is the energy that have 5\% of the yield
of point A. The first graph is the yield of neutrons counted by long
counter detector time normalized, and the energies is the energies
the accelerator measured before it hit the gold foil, which spread
the energy to a wide spectra. The second graph is the same as the
first one after calibration of the energy axis with our new method.
At the third graph the blue graphs is the actual derivation and the
green graph is fit of superposition of two gaussian, which give the
best results for this method. The reference energy is the global maximum
of the fit. Exact values can be seen at table \eqref{tab:real data}.}
\end{figure}

all the data as we extract can be seen at table \eqref{tab:real data}.
\begin{table}[H]
\begin{centering}
\begin{tabular}{|c|c|c|c|c|c|c|}
\hline 
index & energy sigma & low energy & reference energy & energy difference & energy sigma{*} & reference energy value{*}\tabularnewline
\hline 
\hline 
1 & 1.5 & 1871 & 1873.6 & 2.6 & -0.1 & 1882.1\tabularnewline
\hline 
2 & 15 & 2015 & 2070.4 & 55.4 & 24.37 & 1899\tabularnewline
\hline 
3 & 20 & 2020 & 2073.2 & 53.2 & 23.35 & 1898\tabularnewline
\hline 
\end{tabular}
\par\end{centering}
\caption{\label{tab:real data}The data from the real experiment,which extracted
by the same means as at the simulations. Energy's value presented
in keV}
\end{table}

\section{summary}

As we said this method doesn't take in to account any physical dimensions
of the problem. Which make it an independent new way to analysis of
the experimental data.

To use this method do the following: 
\begin{itemize}
\item Fit your data as good as you can.
\item Derive this fit and find the energy that gives the maximum. 
\item Find the neutron yield at that energy. 
\item Find the energy at which you measured 5\% yield.
\item Subtract one from the other and take the absolute value; this value
is $\vartriangle E$.
\item To find your approximate energy distribution use equation \eqref{eq:general}
as shown in equations \eqref{eq:sigma 1} and \eqref{eq:sigma 2}.
\item Change the energy value of the high energy according to equation \eqref{eq:calibration}
and change the axis values with the same scale as before, now using
the new energy as reference point.
\end{itemize}
VERY IMPORTANT - make sure the low energy yield value is more than
5\% !!! else you need to extract the fits for the sigma energy and
the real value of reference energy, referring to the minimum present
you can get from the specific data. our fits is relevant for long
counter detector and thick lithium target.


\begin{thebibliography}{1}
\bibitem{simlit}M. Friedman, G. Feinberg, D. Berkovits, Y. Eisen,
M. Paul, A. Shor, Neutron Spectrum of Liquid-Lithium Target at Soreq
Applied Research Accelerator Facility, to be published

\bibitem[2]{key-1}Log book I page 22, threshold scan target , 16-3-2010

\bibitem[3]{key-2}Log book I page 68, threshold scan target with
gold foil, 24-3-2010

\bibitem[4]{key-3}Log book II page 15, target B1, 
\end{thebibliography}
\end{document}